%% file: moriond.tex
\pdfoutput=1
\documentclass{moriond}
\usepackage{multirow}

\def\be{\begin{equation}}
\def\ee{\end{equation}}
\def\bea{\begin{eqnarray}}
\def\eea{\end{eqnarray}}

\input{definitions.tex}

\begin{document}
\vspace*{2.5cm}
\title{Simulated Effects of $1/f$ Noise on an SKA Intensity Mapping Survey}

\author{Stuart E. Harper, Clive Dickinson, Richard Battye, Lucas Olivari}

\address{Jodrell Bank Centre for Astrophysics, School of Physics and Astronomy, University of Manchester, Alan Turing Building,Manchester M13 9PL, England}

\maketitle\abstracts{
	It has been proposed recently that the SKA1-MID could be used to conduct an HI intensity mapping survey that could rival upcoming Stage IV dark energy surveys. However, measuring the weak HI signal is expected to be very challenging due to contaminations such as residual Galactic emission, RFI, and instrumental $1/f$ noise. Modelling the effects of these contaminants on the cosmological HI signal requires numerical end-to-end simulations. Here we present how $1/f$ noise within the receiver can double the effective uncertainty of an SKA-like survey to HI on large angular scales ($\ell < 50$).
}

\section{Introduction}

It has long been considered that the 21\,cm spin-flip transition of neutral hydrogen (HI) could be used as a tracer of large-scale structure (LSS) throughout the Universe. However, the low surface brightness of HI emission has meant that only a small number of individual galaxies have been surveyed at low redshifts.\cite{Zwaan2005,Martin2010} A way forward for HI cosmology is to use a technique called intensity mapping (IM),\cite{Battye2013} which measures the total integrated HI signal within voxels that contain many individual galaxies. IM is not limited to detecting HI emission from each independent galaxy, but instead treats the HI signal as a continuous field.

The SKA1-MID as an array of single dishes could perform an IM survey with sensitivities to LSS comparable to Euclid-like Stage IV dark energy surveys.\cite{Bull2015,Santos2015} However, the HI signal is weak relative to other foreground emissions making HI emission challenging to isolate.\cite{BigotSazy2015} In particular, the difficulties associated with IM surveys have been recently highlighted by GBT observations.\cite{Wolz2015} Therefore it is likely IM surveys will not be sensitivity limited, but instead limited by systematic effects. Modeling these effects requires end-to-end simulations. Here, we present a preliminary discussion of the effect of $1/f$ noise on the recovery of the observed HI power spectrum for an SKA-like IM survey.

\section{Simulations}\label{sec:simulations}

The setup of the simulated SKA1-MID survey was designed to be similar to that described in Bull et al. 2015.\cite{Bull2015} There are 180 dishes with a system temperature of 44\,K, a frequency range of $350 - 1050$\,MHz, and a frequency resolution of 35\,MHz. The survey full-width half-maximum is 0.8\,degrees. The specific observing strategy is continuous azimuthal rotation at a single fixed elevation of 50\,degrees, and a total observing time of 30\,days.

The simulated observed sky was created by combining maps of the cosmological HI signal, Galactic continuum foregrounds and $1/f$ instrumental noise. 
A \texttt{CAMB}-generated matter power spectrum was used to deduce the projected HI angular power spectra for each redshift slice.\cite{Battye2013} The amplitude of Galactic foreground emission from synchrotron radiation was calculated using the reprocessed Haslam 408\,MHz map,\cite{Remazeilles2015} and spectral indicies measured between 408\,MHz and 2.3\,GHz.\cite{Platania2003} Free-free emission was simulated using a combined all-sky H$\alpha$ map.\cite{Dickinson2003}

\begin{figure}\label{fig:sky}
  \centering
    \includegraphics[height=0.70\textwidth,trim={1cm 0cm 1cm 0cm},clip, angle=270]{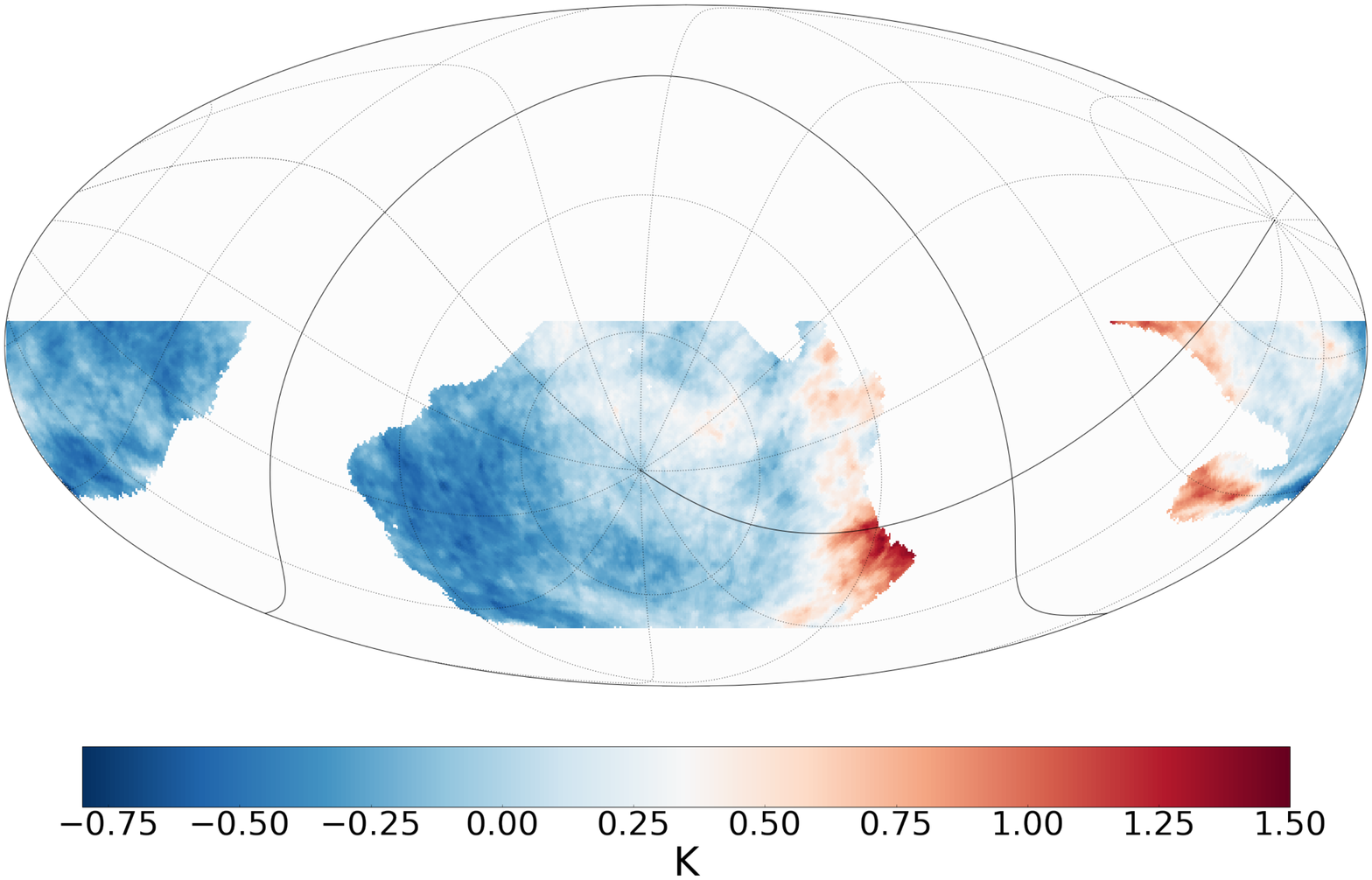}
    \includegraphics[height=0.70\textwidth,trim={1cm 0cm 1cm 0cm},clip, angle=270]{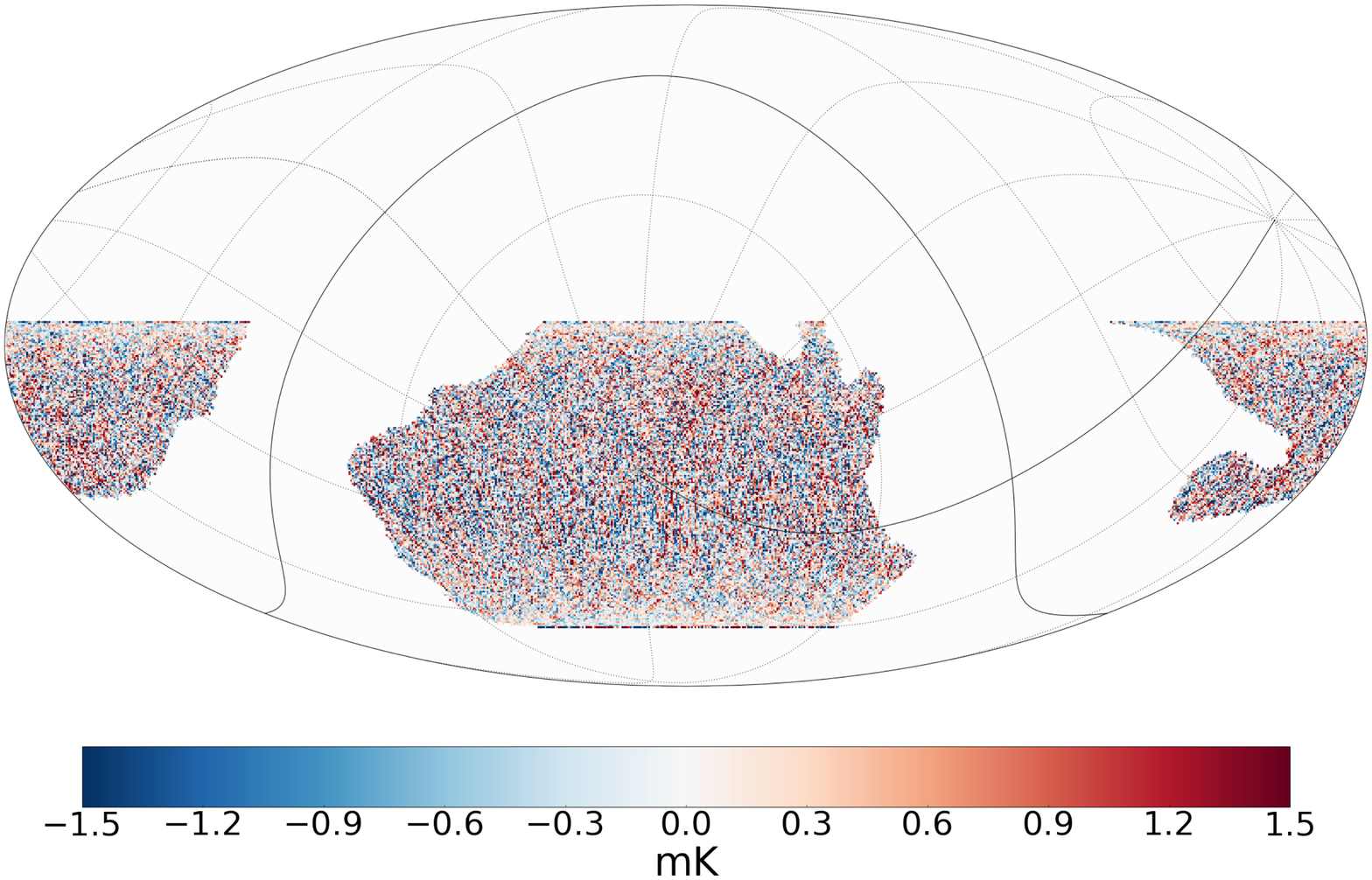}
    \includegraphics[height=0.70\textwidth,trim={1cm 0cm 1cm 0cm},clip, angle=270]{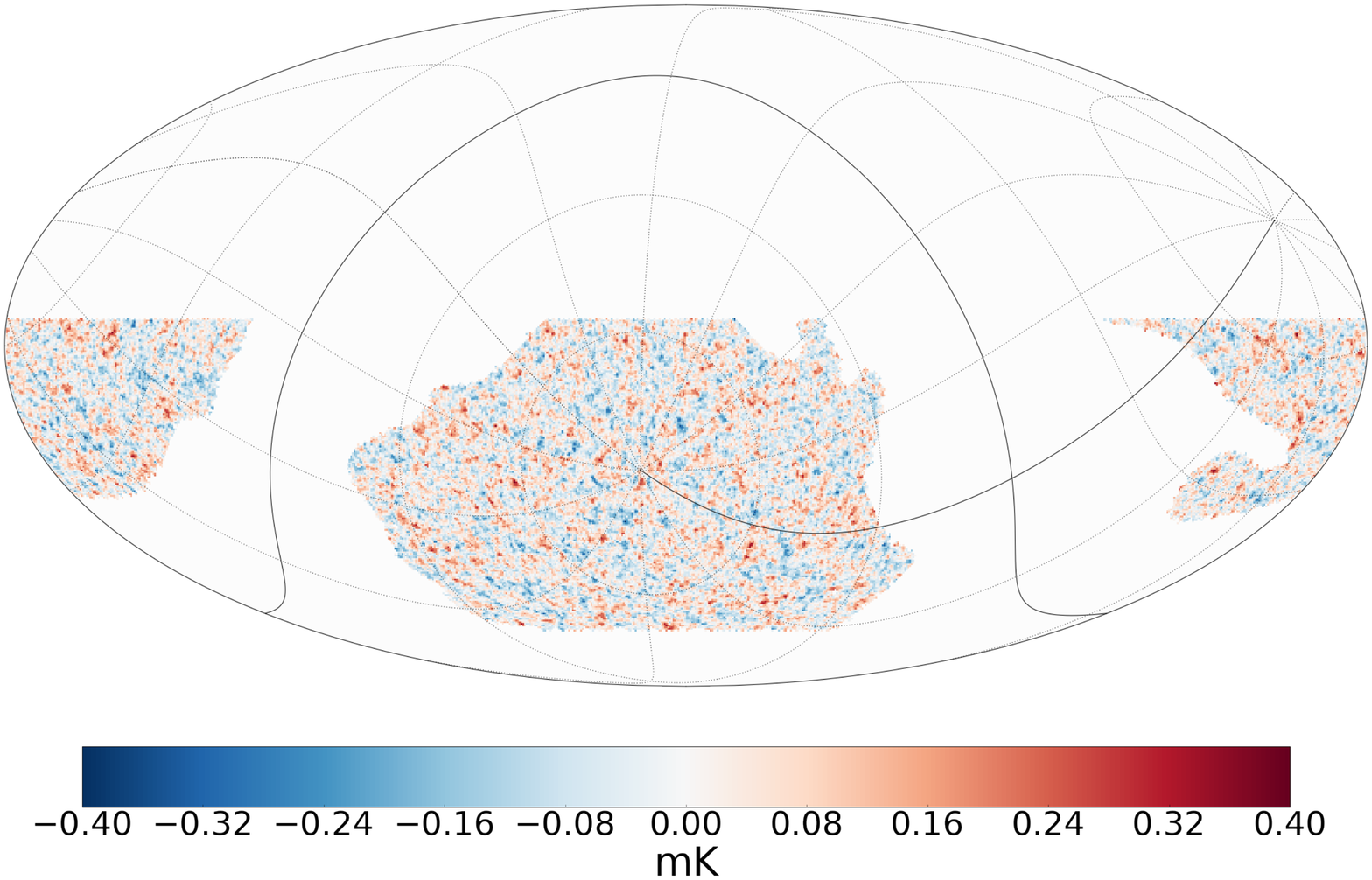}
    \caption{Molleweide projections of the simulated sky at one frequency in the celestial frame overlaid with graticules in the Galactic coordinate frame. Note that the strip appears incomplete due to masking of the Galactic plane. The amplitude of the spatial brightness variations of the foregrounds (\textit{top}, K) and $1/f$ noise (\textit{middle}, mK) here dominate the cosmological HI signal (\textit{bottom}, mK) by orders of magnitude.}
\end{figure}

The $1/f$ noise maps were created from binning time-ordered data (TOD) produced from the Fourier transform of a $1/f$ noise spectrum.\cite{BigotSazy2015,Seiffert2002} 90 noise map realisations were made at knee frequencies ($\nu_{\mathrm{k}}$) 0.1, 1, and 10\,Hz. Figure \ref{fig:sky} shows the survey area coverage and the relative brightnesses of the foregrounds, $1/f$ noise, and HI signal for a specific realisation. The $1/f$ noise model induces more power on longer timescales but these timescales contain little sky information. Therefore, filtering was applied to the TOD on timescales equal to the period of each azimuthal scan ring.

%The 1/f noise was generated by assuming that the voltage response of each SKA receiver had a temporal power spectrum that could be described as\,\cite{Seiffert2002},
%\begin{equation}\label{eqn:fnoise}
%	P(\nu) = \sigma_{w}^2 \left(\frac{\nu_{k}}{\nu}\right)^{\alpha},
%\end{equation}
%where $P(\nu)$ is the power spectrum, $\sigma_{w}$ is the white-noise amplitude, $\nu_{k}$ is the 1/f knee frequency, and $\alpha$ is the 1/f spectral index. 90 noise map realisations were made at knee frequencies 0.1, 1, and 10\,Hz. Figure \ref{fig:sky} shows the survey area coverage, and the relative brightnesses of the foregrounds, 1/f noise, and HI signal.

It was found that $1/f$ noise with a perfectly flat frequency (bandpass) response could be trivially removed using blind component separation methods such as principle component analysis (PCA). However, in reality the frequency response of $1/f$ noise is not expected to be flat. Therefore a toy bandpass model with a 15\,per\,cent smooth variation in gain at low and high frequencies was used to simulate the $1/f$ noise frequency response. Another interesting effect tested by the simulations were two bandpass calibration models. One model assumed perfect calibration with zero uncertainty in frequency space, and the other applied a 5\,per\,cent Gaussian uncertainty to each frequency channel.

\begin{figure}\label{fig:bias}
  \centering
    \includegraphics[height=0.49\textwidth,width=0.25\textwidth,trim={4cm 1cm 4cm 1cm},clip, angle=270]{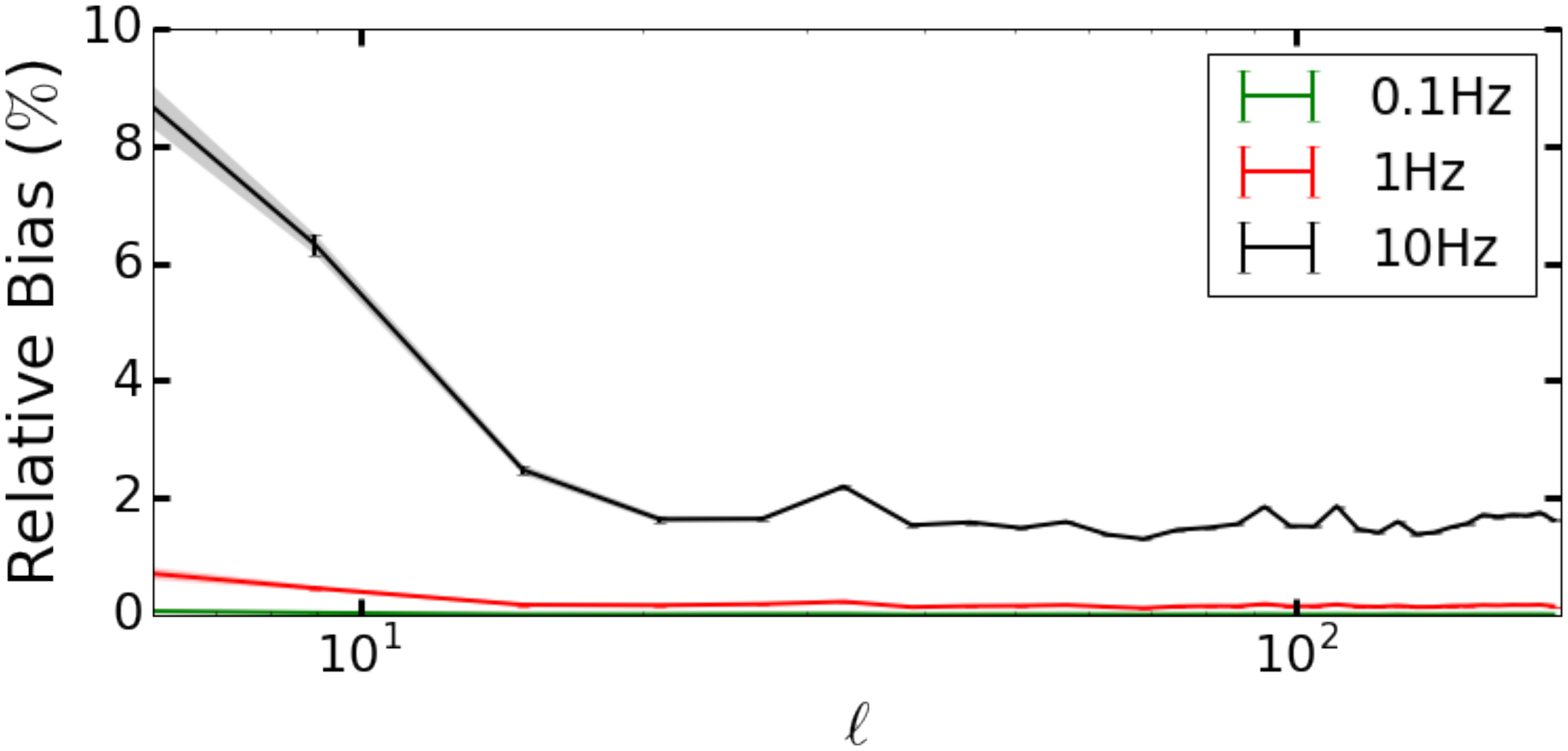}
    \includegraphics[height=0.49\textwidth,width=0.25\textwidth,trim={4cm 1cm 4cm 1cm},clip, angle=270]{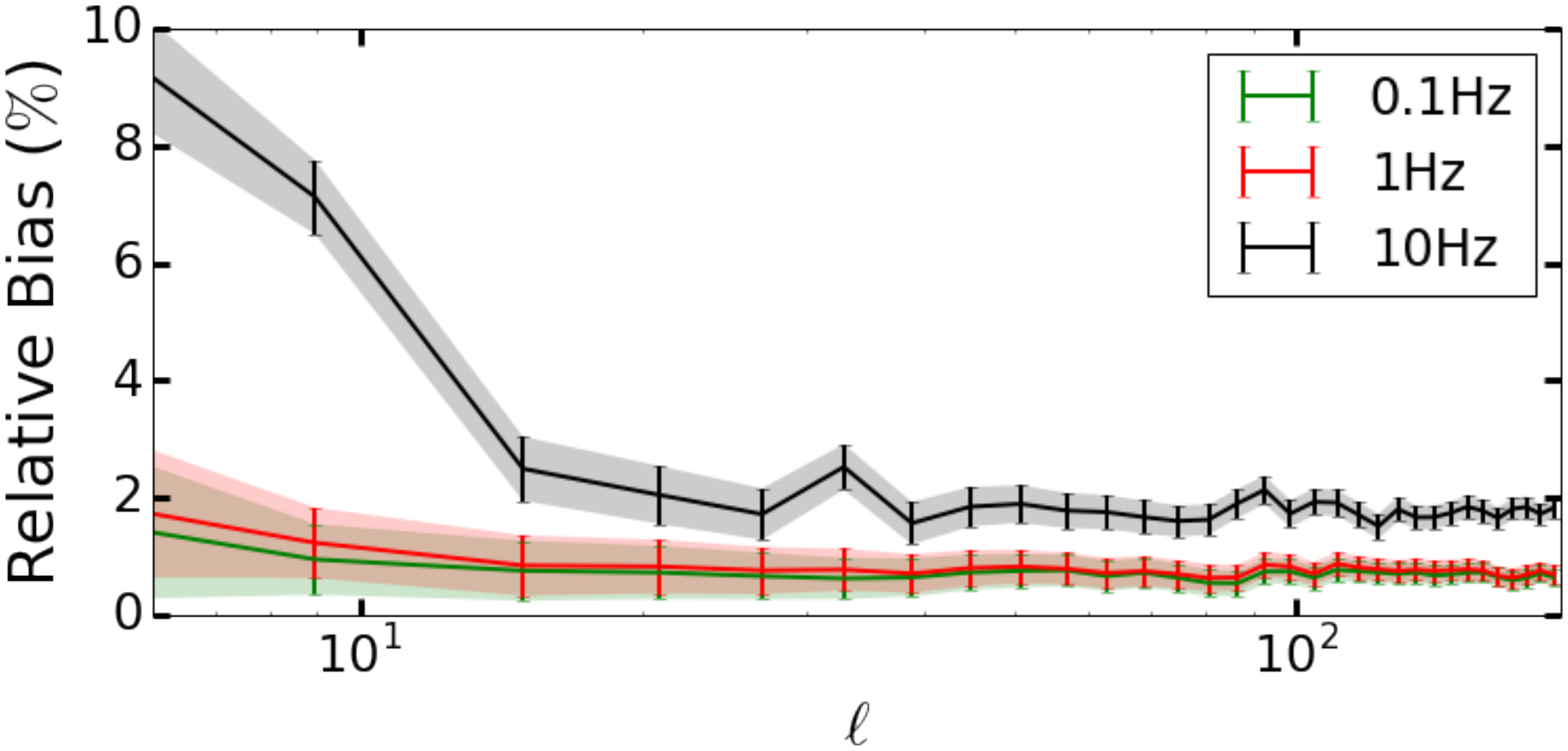}
    \caption{Magnitude of the bias in the estimation of the power spectrum due to $1/f$ noise. \textit{Left}: Perfect calibration across bandpass. \textit{Right}: 5\% Gaussian calibration error across bandpass.}
\end{figure}

The input HI maps were recovered by performing PCA based blind foreground removal. The recovered HI power spectra were then calculated using \texttt{PolSpice}\,\cite{Szapudi2001}, accounting for incomplete sky coverage. Two figures-of-merit were used to determine the robustness of the recovered spectra ($C_{\ell}^{k}$). First, the absolute mean fractional bias in the power spectrum defined as
\small
\begin{equation}\label{eqn:bias}
	b_{\nu} = \left| \left< \frac{C_{\ell}^{k,\nu} - C_{\ell}^{0,\nu} }{C_{\ell}^{\mathrm{HI}} } \right> \right| ,
\end{equation}
\normalsize
where $C_{\ell}^{k,\nu}$ is the recovered power spectrum for a knee frequency $k$, $C_{\ell}^{0,\nu}$ is the recovered power spectrum for data with zero knee frequency, and $C_{\ell}^{\mathrm{HI}}$ is the measured power spectrum of the input HI map. Figure \ref{fig:bias} shows the resulting bias for both perfect and 5\,\% bandpass calibration errors. In both cases the additional bias due to $1/f$ noise is always less than a few per\,cent across all $\ell$ with a slightly higher bias on large-scales, shown also in Table \ref{table:bias}.

\begin{table}[h]
\centering
\caption{The measured bias in $C_{\ell}$ as a percentage due to $1/f$ noise for calibration uncertainties ($\sigma_{\mathrm{cal}}$) of 0 and 5\,per\,cent over three ranges of $\ell$ for three knee frequencies ($\nu_{\mathrm{k}}$) of 0.1, 1, and 10\,Hz.}
\label{table:bias}
\begin{tabular}{|l|c|cc|cc|cc|} \hline
\multicolumn{2}{|l|}{$\ell$}                    & \multicolumn{2}{c|}{$0-10$}     & \multicolumn{2}{c|}{$10-50$}   & \multicolumn{2}{c|}{$50-190$}     \\ \hline
\multicolumn{2}{|l|}{$\sigma_{\mathrm{cal}}$}   & 0\%           & 5\%             & 0\%                            & 5\%            & 0\%            & 5\%   \\ \hline
\multirow{3}{*}{$\nu_{\mathrm{k}}$ (Hz)} & 0.1 & $0.04 $        & $1.0 $          & $0.018 $                       & $0.7 $ & $0.016$ & $0.7$ \\
                                         & 1   & $0.46$         &  $1.2 $         & $0.18 $                        & $0.8 $  & $0.162$ & $0.8$ \\
                                         & 10  & $6.3 $         &  $7.1 $         & $1.85 $                        & $2.0 $  & $1.56$ & $1.8$ \\ \hline
\end{tabular}
\end{table}

\begin{figure}\label{fig:noise}
  \centering
    \includegraphics[height=0.49\textwidth,width=0.25\textwidth,trim={4cm 1cm 4cm 1cm},clip, angle=270]{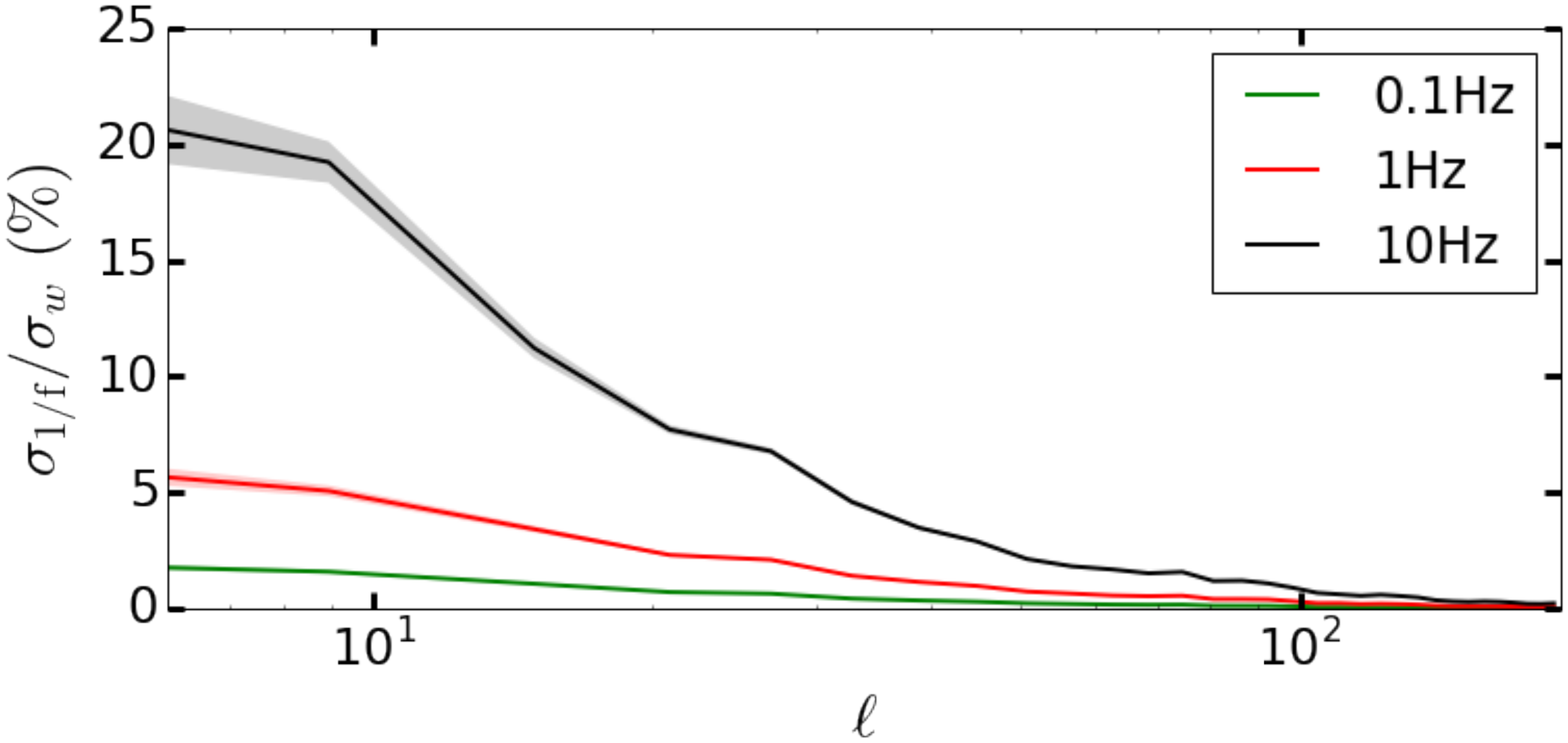}
    \includegraphics[height=0.49\textwidth,width=0.25\textwidth,trim={4cm 1cm 4cm 1cm},clip, angle=270]{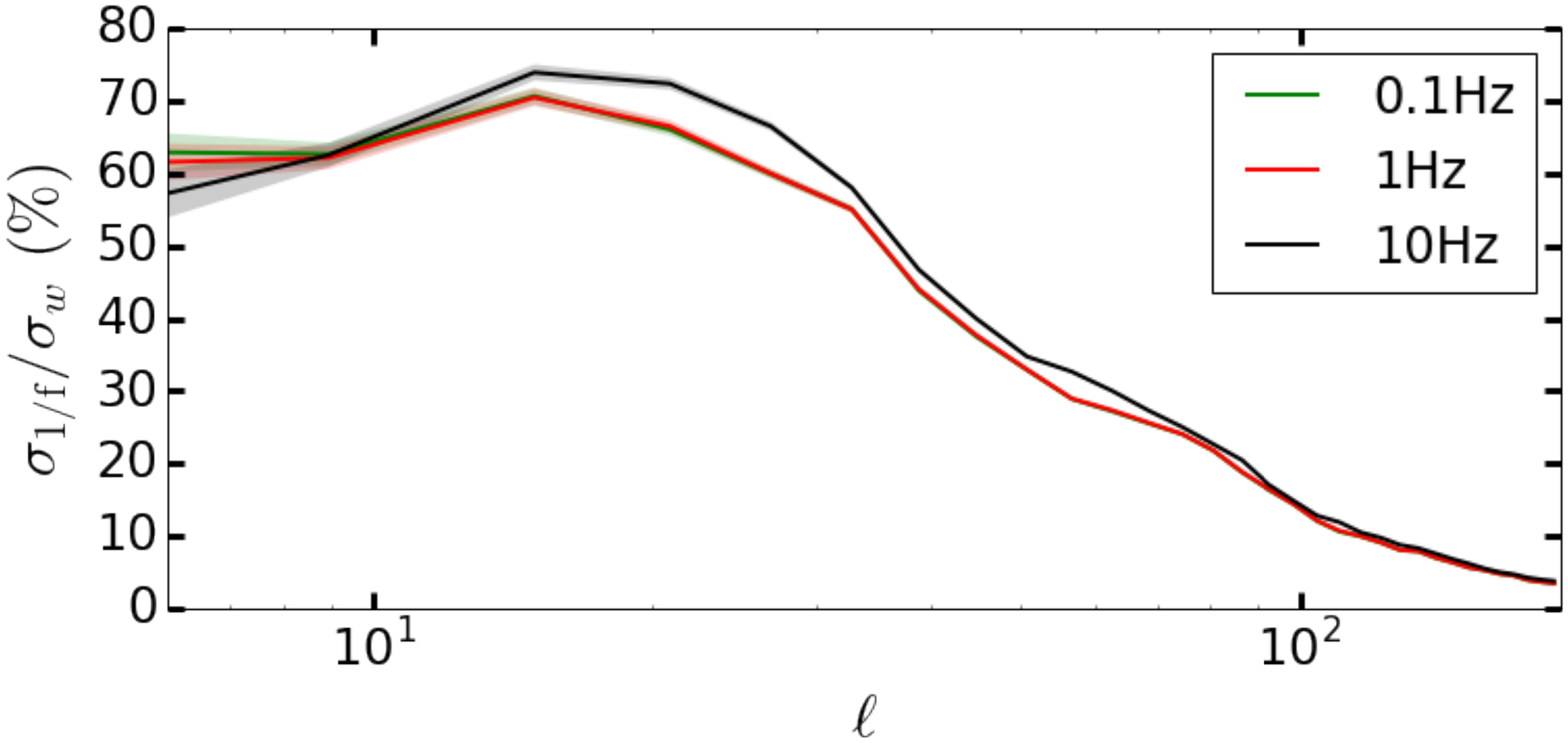}
    \caption{Ratio of the additional noise on each $C_{\ell}$ due to $1/f$ noise over the calculated white-noise from the simulated SKA survey. \textit{Left}: Perfect bandpass calibration. \textit{Right}: 5\% Gaussian calibration error across bandpass.}
\end{figure}

\begin{table}[]
\centering
\caption{The increase in uncertainty on each $C_{\ell}$ as a percentage due to $1/f$ noise over the expected uncertainty from white-noise. For each range of $\ell$ results based on perfect and 5\% calibration uncertainties are presented.}
\label{table:noise}
\begin{tabular}{|l|c|cc|cc|cc|} \hline
\multicolumn{2}{|l|}{$\ell$}                    & \multicolumn{2}{c|}{0-10}    & \multicolumn{2}{c|}{10-50}   & \multicolumn{2}{c|}{50-190}    \\ \hline
\multicolumn{2}{|l|}{$\sigma_{\mathrm{cal}}$}   & 0\%            & 5\%         & 0\%            & 5\%         & 0\%            & 5\%          \\ \hline
\multirow{3}{*}{$\nu_{\mathrm{k}}$ (Hz)} & 0.1 & $1.6  $ & $62  $   & $0.59  $  & $55  $ & $0.090 $ & $13  $ \\
                                         & 1   & $5.1  $ & $62 $   & $1.90  $ & $56  $  & $0.285  $ & $13  $ \\
                                         & 10  & $19.2  $ & $63  $   & $6.1 $ & $60  $  & $0.80  $ & $14  $ \\ \hline
\end{tabular}
\end{table}

The second figure-of-merit is the ratio of the additional uncertainty on each recovered $C_{\ell}$ due to the $1/f$ noise ($\sigma_{c}$) over the expected white-noise level ($\sigma_{w}$) defined as, $r_{\nu} = \sigma_{c}/\sigma_{w}$. Figure \ref{fig:noise} and Table \ref{fig:noise} show the additional uncertainties assuming 0\,\% and 5\,\% frequency calibration errors. The increase at low-$\ell$ when there is no calibration errors is small, less than 20\,per\,cent for all knee frequencies and all $\ell$. However, the $1/f$ noise combined with a calibration error of 5\,\% effectively doubles the noise level of the survey even for the smallest knee frequency of 0.1\,Hz. 

%\b

\section{Discussion}

The results presented in Section \ref{sec:simulations} indicate that just $1/f$ receiver noise combined with a 5\,\% calibration error can almost half the sensitivity of an SKA-like survey to large-scales ($\ell < 50$.). However, the results of Section \ref{sec:simulations} are dependent upon the complexity of the frequency response of the $1/f$ noise. In reality the $1/f$ noise is expected to be more complex in frequency than the toy model assumed here, which could lead to large increases in the bias and uncertainty due to $1/f$ noise. Further, the calibration errors are unlikely to have a Gaussian distribution but will have skewed or correlated errors. For more details see Harper et al. 2016 (\textit{in prep.}). The effects of component separation on the results also require future investigation, as other techniques will yield different results to those of the PCA method. Finally, it should be pointed out that the complexity of real observations, due to other systematics such as standing waves or RFI, will further increase the bias and uncertainty in the HI power spectra. 

\section*{References}
\small
%\begin{thebibliography}{99}
%\end{thebibliography}
\bibliography{moriond}
\end{document}

%% file: definitions.tex
\def\reff@jnl#1{{\rm#1\/}}
\def\aj{\reff@jnl{AJ}}                  % Astronomical Journal 
\def\araa{\reff@jnl{ARA\&A}}            % Annual Review of Astron and Astrophys 
\def\apj{\reff@jnl{ApJ}}                % Astrophysical Journal 
\def\apjl{\reff@jnl{ApJ}}               % Astrophysical Journal, Letters 
\def\aplett{\reff@jnl{ApJ}}             % Astrophysical Journal, Letters 
\def\apjs{\reff@jnl{ApJS}}        % Astrophysical Journal, Supplement 
\def\ao{\reff@jnl{Appl.Optics}}         % Applied Optics 
\def\apss{\reff@jnl{Ap\&SS}}            % Astrophysics and Space Science 
\def\aap{\reff@jnl{A\&A}}               % Astronomy and Astrophysics 
\def\aapr{\reff@jnl{A\&A~Rev.}}  % Astronomy and Astrophysics Reviews 
\def\aaps{\reff@jnl{A\&AS}}    % Astronomy and Astrophysics, Supplement 
\def\azh{\reff@jnl{AZh}}                % Astronomicheskii Zhurnal 
\def\baas{\reff@jnl{BAAS}}              % Bulletin of the AAS 
\def\jrasc{\reff@jnl{JRASC}}            % Journal of the RAS of Canada 
\def\memras{\reff@jnl{MmRAS}}           % Memoirs of the RAS 
\def\mnras{\reff@jnl{MNRAS}}            % Monthly Notices of the RAS 
\def\nar{\reff@jnl{NAR}}           	% National Astronomy Review 
\def\pasa{\reff@jnl{PASA}}			  %Publications of the Astronomical ralia 
\def\pra{\reff@jnl{Phys.Rev.A}}         % Physical Review A: General Physics 
\def\prb{\reff@jnl{Phys.Rev.B}}   % Physical Review B: Solid State 
\def\prc{\reff@jnl{Phys.Rev.C}}         % Physical Review C 
\def\prd{\reff@jnl{Phys.Rev.D}}         % Physical Review D 
\def\prl{\reff@jnl{Phys.Rev.Lett}}      % Physical Review Letters 
\def\pasp{\reff@jnl{PASP}}              % Publications of the ASP 
\def\pasj{\reff@jnl{PASJ}}              % Publications of the ASJ 
\def\qjras{\reff@jnl{QJRAS}}            % Quarterly Journal of the RAS 
\def\skytel{\reff@jnl{S\&T}}            % Sky and Telescope 
\def\solphys{\reff@jnl{Solar~Phys.}}    % Solar Physics 
\def\sovast{\reff@jnl{Soviet~Ast.}}     % Soviet Astronomy 
\def\ssr{\reff@jnl{Space~Sci.Rev.}}     % Space Science Reviews 
\def\zap{\reff@jnl{ZAp}}                % Zeitschrift fuer Astrophysik 
\def\nat{\reff@jnl{Nature}}             % Nature 